\documentstyle[11pt]{article}
\begin{document}
\vspace{20mm}
\title{Yang-Baxter Equation for the R-matrix of 1-D SU(n) Hubbard Model}
\author{Dan-tao Peng$^a$ \thanks{Email:dtpeng@phy.nwu.edu.cn} \hspace{5mm}
           Rui-hong Yue$^{a, b}$ \thanks{Email:yue@phy.nwu.edu.cn}\\[3mm]
           $^a$Institute of Modern Physics, Northwest University\\
           Xi'an, 710069, P. R. China\\
           $^b$Institute of Theoretical Physics, Academica Science\\
           Beijing, 100080, P. R. China}
\date{}
\maketitle
\begin{abstract}
Based on the tetrahedral Zamolodchikov algebra, we prove the Yang-Baxter
equation for the R-matrix of 1-D $SU(n)$ Hubbard model. Furthermore, we
present generalizations of the model.

\vspace{5mm}

\end{abstract}

\bigskip

\setcounter{equation}{0}

\section{Introduction}

\indent

The Hubbard model is one of the significant models in the study of
strongly correlated electronic systems which might reveal an enlightenning
role in understanding the mysteries of the high-$T_C$ superconductivity.
The 1-D Hubbard model also favours a lot of properties of integrable
models in non-pertubative quantum field theory and mathematical physics.
Since Lieb and Wu\cite{Lieb} solved the 1-D Hubbard model by Bethe ansatz
method in 1968, based on their results (Lieb and Wu's Bethe ansatz
equations), many works\cite{AO, GV, CF, FW, AAJ, MT1, HS, FK, MT2, TK,
NTA, HV, JPA} have been done extensively to clarify the physical
properties of this model. Although there were lots of works on the Hubbard
model, the integrability was finished until 1986 by Shastry\cite{Shastry},
Olmedilla and Wadati\cite{OW} in both boson and fermion graded versions.
However, the Yang-Baxter equation for the $R$-matrix of the model obtained
by Shastry  was proved until 1995 by Shiroishi and Wadati in Refs.
\cite{SW1, SW2, SW3} and a generalization of the Shastry's bilayer vertex
model was also presented in \cite{SW1}. Moreover, the eigenvalue of the
transfer matrix related to the Hubbard model was suggested in Ref.
\cite{Shastry} and proved through different methods in Refs. \cite{RT,
Martins2}.

Based on the knowledge of Lie algebra, Maassarani and Mathieu succeeded in
constructing the Hamiltonian of the $SU(n)$ XX model and showed its
integrability\cite{MM}. Considered two coupled $SU(n)$ XX models, by using
Shastry's method, Maassarani constructed the $SU(n)$ Hubbard
model\cite{Ma1} and found the related $R$-matrix which ensures the
integrability of the 1-D $SU(n)$ Hubbard model\cite{Ma2}. (It was also
proved by Martins for $n=3, 4$.\cite{Martins1}, and by Yue and
Sasaki\cite{Yue} for general $n$ in terms of Lax pair formulism.) The
exact Solution of the $SU(3)$ Hubbard model was given in Ref.\cite{HPY}.
But the Yang-Baxter equation for the given $R$-matrix was not proved.

The main purpose of the present paper is to prove the Yang-Baxter equation
for the $R$-matrix of 1-D $SU(n)$ Hubbard model following method
suggested in Ref.\cite{SW1}. In section 2 we review the model and its
integrability. We present the L-operator and the $R$-matrix of the model
and formulate the Yang-Baxter relation. In section 3 we construct the
tetrahedral Zamolodchikov algebra related to the $SU(n)$ Hubbard model.
The Yang-Baxter equation for the cooresponding $R$-matrix was proved in
section 4 and we also present a generalization of the model in this
section. In section 5 we make some conclusion remarks.

\section{The 1-D $SU(n)$ Hubbard model and its integrability}

\indent

The Hamiltonian of the 1-D $SU(n)$ Hubbard model is:
\begin{equation}
\label{Hamiltonian}
H = \sum_{k=1}^L \sum_{\alpha=1}^{n-1}(E^{n \alpha}_{\sigma, k}
    E^{\alpha n}_{\sigma, k+1} + E^{\alpha n}_{\sigma, k}
    E^{n \alpha}_{\sigma, k+1} + E^{n \alpha}_{\tau, k}
    E^{\alpha n}_{\tau, k+1} + E^{\alpha n}_{\tau, k}
    E^{n \alpha}_{\tau, k+1}) + \frac{U n^2}{4}\sum_{k=1}^L
    C^{(\sigma)}_{k}C^{(\tau)}_{k},
\end{equation}
where $U$ is the Coulumb coupling constant, and $E^{\alpha \beta}_{a, k}
(a=\sigma, \tau)$ is a matrix with an one at row $\alpha$ and column
$\beta$ and zeros otherwise:
$$
(E^{\alpha \beta})_{l m}=\delta^{\alpha}_{l}\delta^{\beta}_{m}.
$$
The subscripts $a$ and $k$ stand for two different $E$ operators at
$k$-th site $(k=1, \cdots, L)$. The $n \times n$ diagonal matrix $C$ is
defined by $C = \sum_{\alpha < n}E^{\alpha \alpha} - E^{n n}$. We also
assume the periodic boundary condition: $E^{\alpha \beta}_{k+L} =
E^{\alpha \beta}_{k}$.

In this model, the system has two types of particles named $\sigma$ and
$\tau$ respectively, and each particle can occupy $(n-1)$ possible states.
The same type of particles can not appear in one site, but two different
types of particles can occupy a same site. We denote $|n\rangle_j$ as the
vaccum state of $j$-th site, $|1\rangle_j, |2\rangle_j, \cdots,
|n-1\rangle_j$ as the $(n-1)$ possible one particle states of $j$-th site.
Under the following basis:
$$
|1\rangle_j = \left (\begin{array}{c}
1\\
0\\
\vdots\\
0
\end{array}\right )_j,
|2\rangle_j = \left (\begin{array}{c}
0\\
1\\
\vdots\\
0
\end{array}\right )_j, \cdots,
|n-1\rangle_j = \left (\begin{array}{c}
0\\
\vdots\\
1\\
0
\end{array}\right )_j,
|n\rangle_j = \left (\begin{array}{c}
0\\
\vdots\\
0\\
1
\end{array}\right )_j,
$$
It could be easily proved that: $E^{\alpha n}_j |n\rangle_j =
|\alpha\rangle_j, E^{n \alpha}_j |n\rangle_j = 0, E^{n \alpha}_j
|\alpha\rangle_j = |n\rangle_j, E^{n \alpha}_j |n\rangle_j = 0$. This
means that the operators $E^{\alpha n}$ and $E^{n \alpha}$ can be
interpreted as the particle creation and annihilation operators
respectively. $E^{\alpha n}_j$ can create a $|\alpha\rangle_j$ state
particle over the vaccum state $|n\rangle_j$ of $j$-th site, and $E^{n
\alpha}_j$ annihilate a $|\alpha\rangle_j$ state particle to vaccum
state of $j$-th site.

The $SU(n)$ Hubbard model is constructed by considering two coupled
$SU(n)$ XX model, so the Hamiltonian (\ref{Hamiltonian}) consists of two
$SU(n)$ XX model with an interation term between them. The Hamiltonian of
the $SU(n)$ XX model is:
\begin{equation}
H_{XX} = \sum_{k=1}^L \sum_{\alpha=1}^{n-1}(E^{n \alpha}_{k} 
         E^{\alpha n}_{k+1} + E^{\alpha n}_{k} E^{n \alpha}_{k+1}),
\end{equation}
and the corresponding $R$-matrix is:
\begin{eqnarray}
R(\lambda) & = & a(\lambda)[ E^{n n}\otimes E^{n n} + \sum_{\alpha, \beta
< n} E^{\alpha \beta}\otimes E^{\beta \alpha}]\nonumber\\
& & + b(\lambda)\sum_{\alpha < n}(x E^{n n}\otimes E^{\alpha \alpha} +
x^{-1} E^{\alpha \alpha}\otimes E^{n n})\nonumber\\
& & + c(\lambda)\sum_{\alpha < n}(E^{n \alpha} E^{\alpha n} + E^{\alpha n}
E^{n \alpha}),
\end{eqnarray}
where $x=e^{i\delta}$ and $a(\lambda)=\cos(\lambda),
b(\lambda)=\sin(\lambda), c(\lambda)=1$. The functions $a(\lambda),
b(\lambda), c(\lambda)$ satisfy the free-fermion relation: $a^2(\lambda) +
b^2(\lambda) = c^2(\lambda)$.

The $R$-matrix of the $SU(n)$ XX model satisfies regularity property
$R(0)=P$, unitarity condition $R_{12}(\lambda)R_{21}(-\lambda) =
\cos^2(\lambda)Id$ and Yang-Baxter equation (YBE):
\begin{equation}
R_{31}(\lambda_1)R_{32}(\lambda_2)R_{12}(\lambda_2 - \lambda_1) =
R_{12}(\lambda_2 - \lambda_1)R_{32}(\lambda_2)R_{31}(\lambda_1),
\end{equation}
where $P$ is a permutation operator on the tensor product of two
$n$-dimensional spaces. It is easy to verify that it also satisfies a
decorated Yang-Baxter equation (DYBE):
\begin{equation}
R_{31}(\lambda_1)R_{32}(\lambda_2)C_2 R_{12}(\lambda_2 - \lambda_1) =
R_{12}(\lambda_2 - \lambda_1)C_2 R_{32}(\lambda_2)R_{31}(\lambda_1).
\end{equation}

For the two $SU(n)$ XX model, the $R$-matrix without interaction term is
given by:
\begin{equation}
\bar{R}_{ij}(\lambda) =
R_{ij}^{(\sigma)}(\lambda)R_{ij}^{(\tau)}(\lambda),
\end{equation}
here $R_{ij}^{(\sigma)}(\lambda)$ and $R_{ij}^{(\tau)}(\lambda)$ denote
the $R$-matrices of two $SU(n)$ XX models. Since both
$R_{ij}^{(\sigma)}(\lambda)$ and $R_{ij}^{(\tau)}(\lambda)$ satisfy the
YBE and DYBE, the product $\bar{R}_{ij}(\lambda)$ also satsfy the YBE:
\begin{equation}
\label{YBE}
\bar{R}_{31}(\lambda_1)\bar{R}_{32}(\lambda_2)\bar{R}_{12}(\lambda_2 -
\lambda_ 1) = \bar{R}_{12}(\lambda_2 -
\lambda_1)\bar{R}_{32}(\lambda_2)\bar{R}_{31}(\lambda_1)
\end{equation}
and DYBE:
\begin{equation}
\label{DYBE}
\bar{R}_{31}(\lambda_1)\bar{R}_{32}(\lambda_2)
C^{(\sigma)}_2 C^{(\tau)}_2 \bar{R}_{12}(\lambda_2 - \lambda_ 1) = 
\bar{R}_{12}(\lambda_2 - \lambda_1)C^{(\sigma)}_2 C^{(\tau)}_2 
\bar{R}_{32}(\lambda_2)\bar{R}_{31}(\lambda_1)
\end{equation}
A linear combination of (\ref{YBE}) and (\ref{DYBE}) yields:
\begin{eqnarray}
\label{combination}
& & \bar{R}_{31}(\lambda_1) \bar{R}_{32}(\lambda_2)\left \{ \alpha
\bar{R}_{12}(\lambda_2 - \lambda_1) + \beta C^{(\sigma)}_2 C^{(\tau)}_2
\bar{R}_{12}(\lambda_2 + \lambda_1)\right \} \nonumber\\
& & = \left \{ \alpha \bar{R}_{12}(\lambda_2 - \lambda_1) + \beta 
\bar{R}_{12}(\lambda_2 + \lambda_1) C^{(\sigma)}_2 C^{(\tau)}_2 \right \}
\bar{R}_{32}(\lambda_2) \bar{R}_{31}(\lambda_1),
\end{eqnarray}
here $\alpha$ and $\beta$ are combination coefficients and arbitrary.

For the $SU(n)$ Hubbard model, the two coupled $SU(n)$ XX model, we look
for a solution of the Yang-Baxter
relation (YBR):
\begin{equation}
\label{YBR}
{\cal L}_{31}(\lambda_1){\cal L}_{32}(\lambda_2)R^h_{12}(\lambda_1,
\lambda_2) = R^h_{12}(\lambda_1, \lambda_2){\cal L}_{32}(\lambda_2){\cal
L}_{31}(\lambda_1)
\end{equation}
in the form:
\begin{eqnarray}
&& R^h_{12}(\lambda_1, \lambda_2) = \alpha \bar{R}_{12}(\lambda_2 -
\lambda_1) + \beta \bar{R}_{12}(\lambda_2 + \lambda_1) C^{(\sigma)}_2
C^{(\tau)}_2,\\
&& {\cal L}_{ij}(\lambda) = \bar{R}_{ij}(\lambda) \exp\{h(\lambda)
C^{(\sigma)}_j
C^{(\tau)}_j\}.
\end{eqnarray}
Comparing eq.(\ref{combination}) with the Yang-Baxter relation
(\ref{YBR}), we get a relation:
\begin{equation}
\label{I-relation}
I_1(\lambda_1)I_2(\lambda_2)R^h_{12}(\lambda_1,
\lambda_2)I_1^{-1}(\lambda_1)I_2^{-1}(\lambda_2) =
\alpha \bar{R}_{12}(\lambda_2 - \lambda_1) + \beta C^{(\sigma)}_2
C^{(\tau)}_2 \bar{R}_{12}(\lambda_2 + \lambda_1),
\end{equation}
where
\begin{equation}
I_j(\lambda) = \exp\{h(\lambda)C^{(\sigma)}_j C^{(\tau)}_j\}.
\end{equation}

From (\ref{I-relation}), we have:
\begin{eqnarray}
& & \frac{\beta a(\lambda_2 + \lambda_1) c(\lambda_2 + \lambda_1)}{\alpha
a(\lambda_2 - \lambda_1) c(\lambda_2 - \lambda_1)} = \tanh(h(\lambda_2) -
h(\lambda_1)),\nonumber\\
& & \frac{\beta b(\lambda_2 + \lambda_1) c(\lambda_2 + \lambda_1)}{\alpha
b(\lambda_2 - \lambda_1) c(\lambda_2 - \lambda_1)} = \tanh(h(\lambda_2) +
h(\lambda_1)),
\end{eqnarray}
which give the ratio of $\alpha$ and $\beta$ and constraints on
$h(\lambda_1)$ and $h(\lambda_2)$. The constraints can be written in more
explicit form \cite{Ma2}:
\begin{equation}
\frac{\sinh(h(\lambda_1))}{\sin(2 \lambda_1)} =
\frac{\sinh(h(\lambda_2))}{\sin(2 \lambda_2)} = \frac{n^2 U}{4}.
\end{equation}

Now we have obtained the $R$-matrix of the 1-D $SU(n)$ Hubbard model
\cite{Ma2}:
\begin{eqnarray}
\label{R-matrix}
R^h_{12}(\lambda_1, \lambda_2) = && R^{(\sigma)}_{12}(\lambda_2 -
\lambda_1) R^{(\tau)}_{12}(\lambda_2 - \lambda_1) + \frac{\cos(\lambda_2 
- \lambda_1)}{\cos(\lambda_2 + \lambda_1)}\tanh(h(\lambda_2) -
h(\lambda_1))\nonumber\\
&& \times R^{(\sigma)}_{12}(\lambda_2 + \lambda_1)
R^{(\tau)}_{12}(\lambda_2 + \lambda_1) C^{(\sigma)}_2 C^{(\tau)}_2.
\end{eqnarray}
which satsify the Yang-Baxter relation (\ref{YBR}). This $R$-matrix
depends not only on the diference of the spectral parameters $\lambda_2 -
\lambda_1$, but also on the sum of the spectral parameters $\lambda_2 +
\lambda_1$. This non-additive property allows us to generalize the
Hamiltonian of the 1-D $SU(n)$ Hubbard model (see section 4). 

The monodromy matrix of the model can be defined as:
\begin{equation}
{\cal T}_a(\lambda) = {\cal L}_{La}(\lambda){\cal L}_{L-1 a}(\lambda)
\cdots {\cal L}_{1 a}(\lambda).
\end{equation}
From the Yang-Baxter relation (\ref{YBR}) we know the monodromy matrix
satisfies the
global Yang-Baxter relation:
\begin{equation}
\label{GYBR}
{\cal T}_1(\lambda_1){\cal T}_2(\lambda_2)R^h_{12}(\lambda_1, \lambda_2) =
R^h_{12}(\lambda_1, \lambda_2){\cal T}_2(\lambda_2){\cal T}_1(\lambda_1).
\end{equation}

The corresponding transfer matrix is defined by:
\begin{equation}
\tau(\lambda) = tr_a[{\cal T}_a(\lambda)]
\end{equation}
and from (\ref{GYBR}) it can be easily proved the existence of a commuting
family of transfer matrices
\begin{equation}
[\tau(\lambda_1), \tau(\lambda_2)] = 0.
\end{equation}
Then the integrability of the model was proved.

Using the relation $h(0) = 0$ and $h^\prime(0) = \frac{n^2 U}{4}$ we can
obtain the Hamiltonian of the 1-D $SU(n)$ Hubbard model
(\ref{Hamiltonian}):
\begin{eqnarray}
H & = & \frac{d}{d\lambda}\ln \tau(\lambda)|_{\lambda = 0} = \tau^{-1}(0)
\frac{d}{d\lambda}\tau(\lambda)|_{\lambda = 0}\nonumber\\
& = & \sum_{k=1}^L \sum_{\alpha=1}^{n-1}(E^{n \alpha}_{\sigma, k}
    E^{\alpha n}_{\sigma, k+1} + E^{\alpha n}_{\sigma, k}
    E^{n \alpha}_{\sigma, k+1} + E^{n \alpha}_{\tau, k}
    E^{\alpha n}_{\tau, k+1} + E^{\alpha n}_{\tau, k}
    E^{n \alpha}_{\tau, k+1})\nonumber\\
& & + \frac{U n^2}{4}\sum_{k=1}^L C^{(\sigma)}_{k}C^{(\tau)}_{k}.
\end{eqnarray}

\section{Tetrahedral Zamolodchikov algebra}

\indent

In the above section we have shown the integrability of the 1-D $SU(n)$
Hubbard model. It is natrual to expect that the $R$-matrix
(\ref{R-matrix}) itself satisfy the Yang-Baxter equation(YBE):
\begin{equation}
\label{RYBE}
R^h_{31}(\lambda_3, \lambda_1)R^h_{32}(\lambda_3,
\lambda_1)R^h_{12}(\lambda_1, \lambda_1) = R^h_{12}(\lambda_1,
\lambda_2)R^h_{32}(\lambda_3, \lambda_2)R^h_{31}(\lambda_3, \lambda_1).
\end{equation}

In the $SU(2)$ case, the YBE of the $R$-matrix was proved in 
ref.\cite{SW1} by using the tetrahedral Zemolodchikov algebra (TZA)
\cite{IK, SW3}. In this section, we construct the TZA related to the
$SU(n)$ Hubbard model.

The TZA is defined by the following set of relations:
\begin{equation}
\label{TZA}
{\cal L}^a_{12}{\cal L}^b_{32}{\cal L}^c_{31} = \sum_{def}S^{abc}_{def}
{\cal L}^f_{31}{\cal L}^e_{32}{\cal L}^d_{12}.
\end{equation}
where $a, b, \cdots, f = 0, 1$ and $S^{abc}_{def}$ are some scalar
coefficients.

We take ${\cal L}^0_{jk}$ and ${\cal L}^1_{jk}$ as follows:
\begin{equation}
\label{L}
{\cal L}^0_{jk} = R_{jk}(\lambda_k - \lambda_j), \hspace{1cm}
{\cal L}^1_{jk} = R_{jk}(\lambda_k + \lambda_j)C_k,
\end{equation}
where $R_{jk}(\lambda)$ is the $R$-matrix of the $SU(n)$ XX model as
before. Then we could find the following relations which give the TZA
(\ref{TZA}):
\begin{eqnarray}
\label{S000-S011}
&&{\cal L}^0_{12}{\cal L}^0_{32}{\cal L}^0_{31} = 
{\cal L}^0_{31}{\cal L}^0_{32}{\cal L}^0_{12}, \hspace{1cm}
{\cal L}^0_{12}{\cal L}^1_{32}{\cal L}^1_{31} = 
{\cal L}^1_{31}{\cal L}^1_{32}{\cal L}^0_{12}, \\
\label{S110-S101}
&&{\cal L}^1_{12}{\cal L}^1_{32}{\cal L}^0_{31} = 
{\cal L}^0_{31}{\cal L}^1_{32}{\cal L}^1_{12}, \hspace{1cm}
{\cal L}^1_{12}{\cal L}^0_{32}{\cal L}^1_{31} = 
{\cal L}^1_{31}{\cal L}^0_{32}{\cal L}^1_{12}, \\
\label{S111}
&&{\cal L}^1_{12}{\cal L}^1_{32}{\cal L}^1_{31} = S^{111}_{001}
{\cal L}^1_{31}{\cal L}^0_{32}{\cal L}^0_{12} + S^{111}_{010}
{\cal L}^1_{31}{\cal L}^0_{32}{\cal L}^1_{12} + S^{111}_{100}
{\cal L}^0_{31}{\cal L}^0_{32}{\cal L}^1_{12}, \\
\label{S001}
&&{\cal L}^0_{12}{\cal L}^0_{32}{\cal L}^1_{31} = S^{001}_{111}
{\cal L}^1_{31}{\cal L}^1_{32}{\cal L}^1_{12} + S^{001}_{100}
{\cal L}^0_{31}{\cal L}^0_{32}{\cal L}^1_{12} + S^{001}_{010}
{\cal L}^0_{31}{\cal L}^1_{32}{\cal L}^0_{12}, \\
\label{S010}
&&{\cal L}^0_{12}{\cal L}^1_{32}{\cal L}^0_{31} = S^{010}_{111}
{\cal L}^1_{31}{\cal L}^1_{32}{\cal L}^1_{12} + S^{010}_{100}
{\cal L}^0_{31}{\cal L}^0_{32}{\cal L}^1_{12} + S^{010}_{001}
{\cal L}^1_{31}{\cal L}^0_{32}{\cal L}^0_{12}, \\
\label{S100}
&&{\cal L}^1_{12}{\cal L}^0_{32}{\cal L}^0_{31} = S^{100}_{111}
{\cal L}^1_{31}{\cal L}^1_{32}{\cal L}^1_{12} + S^{100}_{010}
{\cal L}^0_{31}{\cal L}^1_{32}{\cal L}^0_{12} + S^{100}_{001}
{\cal L}^1_{31}{\cal L}^0_{32}{\cal L}^0_{12},
\end{eqnarray}
where the coefficients $S^{abc}_{def}$ are given by
\begin{eqnarray}
&&S^{111}_{001} = \frac{\sin(\lambda_2 + \lambda_1)\cos(\lambda_2 + 
\lambda_3)}{\cos(\lambda_2 - \lambda_1)\sin(\lambda_2 - \lambda_3)},
\hspace{1cm}
S^{111}_{010} = - \frac{\sin(\lambda_2 + \lambda_1)\sin(\lambda_1 + 
\lambda_3)}{\cos(\lambda_2 - \lambda_1)\cos(\lambda_1 - \lambda_3)},
\nonumber\\
&&S^{111}_{100} = - \frac{\sin(\lambda_1 + \lambda_3)\cos(\lambda_2 + 
\lambda_3)}{\cos(\lambda_1 - \lambda_3)\sin(\lambda_2 - \lambda_3)},
\hspace{1cm}
S^{001}_{111} = \frac{\sin(\lambda_2 - \lambda_1)\cos(\lambda_2 - 
\lambda_3)}{\cos(\lambda_2 + \lambda_1)\sin(\lambda_2 + \lambda_3)},
\nonumber\\
&&S^{001}_{100} = \frac{\sin(\lambda_2 - \lambda_1)\sin(\lambda_1 + 
\lambda_3)}{\cos(\lambda_2 + \lambda_1)\cos(\lambda_1 - \lambda_3)},
\hspace{1cm}
S^{001}_{010} = \frac{\sin(\lambda_1 + \lambda_3)\cos(\lambda_2 - 
\lambda_3)}{\cos(\lambda_1 - \lambda_3)\cos(\lambda_2 + \lambda_3)},
\nonumber\\
&&S^{010}_{111} = \frac{\sin(\lambda_2 - \lambda_1)\sin(\lambda_1 - 
\lambda_3)}{\cos(\lambda_2 + \lambda_1)\cos(\lambda_1 + \lambda_3)},
\hspace{1cm}
S^{010}_{100} = \frac{\sin(\lambda_2 - \lambda_1)\cos(\lambda_2 + 
\lambda_3)}{\cos(\lambda_2 + \lambda_1)\sin(\lambda_2 - \lambda_3)},
\nonumber\\
&&S^{010}_{001} = \frac{\sin(\lambda_1 - \lambda_3)\cos(\lambda_2 + 
\lambda_3)}{\cos(\lambda_1 + \lambda_3)\sin(\lambda_2 - \lambda_3)},
\hspace{1cm}
S^{100}_{111} = - \frac{\sin(\lambda_1 - \lambda_3)\cos(\lambda_2 - 
\lambda_3)}{\cos(\lambda_1 + \lambda_3)\sin(\lambda_2 + \lambda_3)},
\nonumber\\
&&S^{100}_{010} = \frac{\sin(\lambda_2 + \lambda_1)\cos(\lambda_2 - 
\lambda_3)}{\cos(\lambda_2 - \lambda_1)\sin(\lambda_2 + \lambda_3)},
\hspace{1cm}
S^{100}_{001} = - \frac{\sin(\lambda_2 + \lambda_1)\sin(\lambda_1 - 
\lambda_3)}{\cos(\lambda_2 - \lambda_1)\cos(\lambda_1 + \lambda_3)},
\end{eqnarray}
Eq.(\ref{S000-S011}) and eq.(\ref{S110-S101}) are equivalent to the YBE
(\ref{YBE}) and DYBE (\ref{DYBE}) respectively. In this sense, the TZA
(\ref{TZA}) can be regarded as a generalization of the YBE and DYBE.

It is important to notice that the products ${\cal L}^a_{12}
{\cal L}^b_{32}{\cal L}^c_{31}$ are not linearly independent as operators
acting on $V_1\otimes V_2 \otimes V_3$ and they satisfy the
following relations:
\begin{eqnarray}
\label{L000}
{\cal L}^0_{12}{\cal L}^0_{32}{\cal L}^0_{31} = 
x_0{\cal L}^0_{12}{\cal L}^1_{32}{\cal L}^1_{31} +
y_0{\cal L}^1_{12}{\cal L}^0_{32}{\cal L}^1_{31} +
z_0{\cal L}^1_{12}{\cal L}^1_{32}{\cal L}^0_{31}, \\
\label{L111}
{\cal L}^1_{12}{\cal L}^1_{32}{\cal L}^1_{31} = 
x_1{\cal L}^1_{12}{\cal L}^0_{32}{\cal L}^0_{31} +
y_1{\cal L}^0_{12}{\cal L}^1_{32}{\cal L}^0_{31} +
z_1{\cal L}^0_{12}{\cal L}^0_{32}{\cal L}^1_{31},
\end{eqnarray}
with
\begin{eqnarray}
&&x_0 = - \frac{\cos(\lambda_1 - \lambda_3)\sin(\lambda_2 - \lambda_3)}
{\cos(\lambda_1 + \lambda_3)\sin(\lambda_2 + \lambda_3)},\hspace{1cm}
y_0 = \frac{\cos(\lambda_2 - \lambda_1)\cos(\lambda_1 - \lambda_3)}
{\cos(\lambda_2 + \lambda_1)\cos(\lambda_1 + \lambda_3)},\\
&&z_0 = \frac{\cos(\lambda_2 - \lambda_1)\sin(\lambda_2 - \lambda_3)}
{\cos(\lambda_2 + \lambda_1)\sin(\lambda_2 + \lambda_3)},\hspace{1cm}
x_1 = - \frac{\cos(\lambda_1 + \lambda_3)\sin(\lambda_2 + \lambda_3)}
{\cos(\lambda_1 - \lambda_3)\sin(\lambda_2 - \lambda_3)},\\
&&y_1 = \frac{\cos(\lambda_2 + \lambda_1)\cos(\lambda_1 + \lambda_3)}
{\cos(\lambda_2 - \lambda_1)\cos(\lambda_1 - \lambda_3)},\hspace{1cm}
z_1 = \frac{\sin(\lambda_2 + \lambda_3)\cos(\lambda_2 + \lambda_1)}
{\sin(\lambda_2 - \lambda_3)\cos(\lambda_2 - \lambda_1)},
\end{eqnarray}

From these relations we know that the linear space spanned by the products
${\cal L}^a_{12}{\cal L}^b_{32}{\cal L}^c_{31}$ is 6-dimensional.

\section{The Yang-Baxter equation for the $R$-matrix of the $SU(n)$
Hubbard model}

\indent

In this section we prove the Yang-Baxter equation for the $R$-matrix of
the 1-D $SU(n)$ Hubbard model (\ref{RYBE}).

Taking into account the form of the $R$-matrix (\ref{R-matrix}), we look
for a solution of the YBE (\ref{RYBE}) in the following form:
\begin{eqnarray}
\label{R-assume}
R^h_{jk}(\lambda_j, \lambda_k) & = & R^{(\sigma)}_{jk}(\lambda_k -
\lambda_j)R^{(\tau)}_{jk}(\lambda_k - \lambda_j) +
\alpha_{jk}R{(\sigma)}_{jk}(\lambda_k + \lambda_j)C^{(\sigma)}_k
R^{(\tau)}_{jk}(\lambda_k + \lambda_j)C^{(\tau)}_k \nonumber\\
& = & {\cal L}^{0 (\sigma)}_{jk}{\cal L}^{0 (\tau)}_{jk} + \alpha_{jk}
{\cal L}^{1 (\sigma)}_{jk}{\cal L}^{1 (\tau)}_{jk}, 
\end{eqnarray}
where ${\cal L}^0_{jk}$ and ${\cal L}^1_{jk}$ have been defined in 
(\ref{L}). If $\alpha_{jk} = 0$, the $R$-matrix satisfies the YBE
(\ref{RYBE}) in a trivial way. Now we look for a non-trivial solution.
Subsitituting the expression (\ref{R-assume}) into the Yang-Baxter
equation (\ref{RYBE}), by means of the tetralhedral Zamolodchikov algebra
and relations (\ref{L000}) and (\ref{L111}), we could find that
$\alpha_{jk}$ must satisfy the following conditions:
\begin{eqnarray}
\label{alpha-condition}
&& \alpha_{12}\sin 2(\lambda_1 + \lambda_2) + \alpha_{31}\sin 2(\lambda_1
+ \lambda_3) = \alpha_{32}\sin 2(\lambda_2 + \lambda_3)\nonumber\\
&& = \frac{1}{\alpha_{31}}\sin 2(\lambda_3 - \lambda_1) +
\frac{1}{\alpha_{12}}\sin 2(\lambda_2 - \lambda_1).
\end{eqnarray}
If we take
\begin{equation}
\alpha_{jk} = \frac{\cos(\lambda_k - \lambda_j)}{\cos(\lambda_k +
\lambda_j)}\tanh(h(\lambda_k) - h(\lambda_j)),
\end{equation}
and impose the constrains
\begin{equation}
\frac{\sinh(h(\lambda_j)}{\sin(2\lambda_j)} = \frac{n^2 U}{4},
\hspace{1cm} (j = 1, 2, 3),
\end{equation}
then the conditions (\ref{alpha-condition}) are satisfied. This proves the
Yang-Baxter equaiton for the $R$-matrix of the 1-D $SU(n)$ Hubbard model.

Besides the YBE (\ref{RYBE}), the $R$-matrix (\ref{R-matrix}) has the
following properties:
\begin{eqnarray}
&& R^{h}_{jk}(0, \lambda) = \frac{1}{\cosh(h(\lambda))}
{\cal L}_{jk}(\lambda),\\
&& R^{h}_{jk}(\lambda_0, \lambda_0) = {\cal P}_{jk},\\
&& R^{h}_{jk}(\lambda_j, \lambda_k)R^{h}_{kj}(\lambda_k, \lambda_j) =
\rho(\lambda_j, \lambda_k) Id,
\end{eqnarray}
where
\begin{equation}
\rho(\lambda_j, \lambda_k) = \cos^2(\lambda_k - \lambda_j)\left \{
\cos^2(\lambda_k - \lambda_j) - \tanh^2(h(\lambda_k) - h(\lambda_j))\right
\},
\end{equation}
and the permutation operator is defined as
\begin{equation}
{\cal P}_{jk} = {\cal P}^{(\sigma)}_{jk} {\cal P}^{(\tau)}_{jk}.
\end{equation}

The Yang-Baxter equation (\ref{RYBE}) implies a more general inhomogeneous
model as:
\begin{equation}
{\cal T}_a(\lambda, \{\lambda_j\}) = R^h_{L a}(\lambda, \lambda_N)
R^h_{L-1 a}(\lambda, \lambda_{L-1}) \cdots R^h_{1 a}(\lambda, \lambda_1),
\end{equation}
where $\lambda_j, (j=1, 2, \cdots, L)$ are the inhomogeneous parameters
obeying the constraints
\begin{equation}
\frac{\sinh(2h(\lambda_j))}{\sin(2\lambda_j)} = \frac{n^2 U}{4},
\hspace{1cm} (j=1, 2, \cdots, L).
\end{equation}
From the Yang-Baxter equation (\ref{RYBE}), we can obtain the global
Yang-Baxter relation:
\begin{equation}
{\cal T}_1(\lambda, \{\lambda_j\}){\cal T}_2(\mu,
\{\lambda_j\})R^h_{12}(\lambda, \mu) = R^h_{12}(\lambda, \mu)
{\cal T}_2(\mu, \{\lambda_j\}){\cal T}_1(\lambda, \{\lambda_j\}),
\end{equation}
which leads to the commutativity
\begin{equation}
[\tau(\lambda, \{\lambda_j\}), \tau(\mu, \{\lambda_j\})] = 0,
\end{equation}
where $\tau(\lambda, \{\lambda_j\})$ is the transfer matrix of the model
\begin{equation}
\tau(\lambda, \{\lambda_j\}) = tr_a {\cal T}_a(\lambda, \{\lambda_j\}).
\end{equation}
The corresponding Hamiltonian is defined as the logarithmic derivative of
the transfter matrix under all inhomogeneous parameters $\lambda_j =
\lambda_0, (j = 1, 2, \cdots, L)$:
\begin{eqnarray}
\label{new-Hamiltonian}
H_{\lambda_0} & = & \frac{d}{d \lambda}\ln \tau(\lambda,
\{\lambda_j = \lambda_0\})|_{\lambda = \lambda_0} = \tau^{-1}(\lambda_0,
\{\lambda_j = \lambda_0\})\frac{d}{d \lambda}\tau(\lambda, \{\lambda_j =
\lambda_0\})\nonumber\\
& = & \sum_{j=1}^L \sum_{\alpha < n}(
E^{n \alpha}_{\sigma j}E^{\alpha n}_{\sigma j+1} +
E^{\alpha n}_{\sigma j}E^{n \alpha}_{\sigma j+1} +
E^{n \alpha}_{\tau j}E^{\alpha n}_{\tau j+1} +
E^{\alpha n}_{\tau j}E^{n \alpha}_{\tau j+1})\nonumber\\
& & + \frac{n^2 U}{4\cosh(2h(\lambda_0))}\sum_{j=1}^L 
B^{(\sigma)}_{j j+1}B^{(\tau)}_{j j+1},
\end{eqnarray}
where
\begin{eqnarray}
B_{j j+1} & = & \cos(2 \lambda_0)(- E^{n n}_j E^{n n}_{j+1} +
\sum_{\alpha, \beta < n} E^{\alpha \alpha}_j E^{\beta
\beta}_{j+1}) + \sin(2 \lambda_0)\sum_{\alpha < n}(E^{n \alpha}_j
E^{\alpha n}_{j+1} + E^{\alpha n}_j E^{n \alpha}_{j+1}) \nonumber \\
& & + \sum_{\alpha < n}(- E^{n n}_j E^{\alpha \alpha}_{j+1} +
E^{\alpha \alpha}_j E^{n n}_{j+1}).
\end{eqnarray}
The arbitrariness of the parameter $\lambda_0$ comes from the non-additive
property of the spectral parameters. If we take $\lambda_0 = 0$, this new
Hamiltonian reduces to (\ref{Hamiltonian}).

Thus, we have obtained a new 1-D $SU(n)$ Hubbard by the Yang-Baxter
equation of the $R$-matrix (\ref{RYBE}).

\section{Conclusions}

\indent

In this paper we have proved the $R$-matrix of the 1-D $SU(n)$ Hubbard
model satisfying the Yang-Baxter equation. We notice that the tetrahedral
Zamolodchikov algebra play an essential role in the proof.

In most lattice systems, the existence of $R$-matrix ensures the
integrability and the $R$-matrix is isomorphic to the $L$-operator. Thus,
the Yang-Baxter equation is a consequence of the Yang-Baxter relation
$R_{12} L_1 L_2 = L_2 L_1 R_{12}$. But, for Hubbard model, the situation
is quite different. The $R$-matrix can not be obtained from $L$-operator,
even if we limit to $SU(2)$ Hubbard model\cite{Shastry, OW}. The
$R$-matrix satisfying the Yang-Baxter equation together with $L$-operator
constitute the complete proof of the integrability.

For $SU(n)$ Hubbard model, the $R$-matrix does not isomorphic to the
$L$-operator. This provides a method to construct a new kind of integrable
system by considering the $R$-matrix as a $L$-operator (fundamental
representation of same algebra). The general representation can be
obtained by fusing the multi fundamental rep. ($R$-matrix). Therefore, one
can get the full representation of the algebra in principle.

In the present paper we have derived out a new Hamitonian
(\ref{new-Hamiltonian}) from the $R$-matrix. The last term introduces a
new kind of interaction. This Hamiltonian is quite different from the
original one (\ref{Hamiltonian}). It also renders a question how to
find the eigenvalue of this Hamiltonian (\ref{new-Hamiltonian}). We will
consider it late.

In the derivation of eq. (\ref{new-Hamiltonian}), we have assumed all the
parameters $\lambda_j$ are same $\lambda_0$. This is not necessary. The
different choice will give out different Hamiltinian. On the other hand,
one can consider the $R$-matrix as a Boltzmann weight in statistical
mechanics. This provides a inhomegeneous lattice statistical medel. The
partition function could be derived out in a similar way.

\end{document}